\documentclass[onefignum,onetabnum]{siamart250211}

\usepackage{graphicx}%
\usepackage{subcaption}%
\usepackage[percent]{overpic}%
\usepackage{multirow}%
\usepackage{amsmath,amssymb,amsfonts}%
\usepackage{mathrsfs}%
\usepackage[title]{appendix}%
\usepackage{xcolor}%
\usepackage{textcomp}%
\usepackage{manyfoot}%
\usepackage{booktabs}%
\usepackage{algorithm}%
\usepackage{algorithmicx}%
\usepackage{algpseudocode}%
\usepackage{listings}%
\usepackage[normalem]{ulem} 
\usepackage{tikz}

\newtheorem{claim}{Claim.}

\newcommand{\depsilon}{\dot{\varepsilon}}
\newcommand{\R}{\mathbb{R}}
\def\*#1{{\mathbf{#1}}} 
\newcommand{\pder}[2]{\dfrac{\partial #1}{\partial #2}}

\begin{document}

\title{On the dynamical stability of skeletal muscle}


\author{
	{Javier A. Almonacid \thanks{Department of Mathematics,Simon Fraser University, Burnaby, Canada.  \email{javiera@sfu.ca}} }
	\and
	{Nilima Nigam \thanks{Department of Mathematics,Simon Fraser University, Burnaby, Canada. \email{nigam@math.sfu.ca}}}		
	\and{
		James M.Wakeling\thanks{Department of Biomedical Physiology and Kinesiology, Simon Fraser University, Burnaby, Canada. \email{wakeling@sfu.ca}}}
}



\maketitle
\begin{abstract}
    There has been debate for over 70-years about whether active skeletal muscle is dynamically stable at lengths greater than its optimal length. The stability of computational muscle models is a critical issue, as it directly affects our ability to simulate muscle deformation across different operating lengths, especially at lengths where muscles are known to remain functional despite model-predicted instabilities.
    In this study, we revisit the question of dynamical stability of ODE-based models of skeletal muscle. 
    In particular, 
    we investigate whether activation-independent tissue properties can provide stability to contractions along the dip region of the total force-length curve. 
    First, using a combination of analytical tools (eigenvalue analysis and non-dimensionalization) and numerical simulations, we confirm that traditional Hill-type muscle models can display divergent dynamics in this region. 
    Then, we propose a stabilized version of a 1D Hill-type muscle model that incorporates the 3D nature of skeletal muscle deformation. This results in a completely convex force-length relationship that can bring robustness to numerical simulations, while preserving the computational efficiency of 1D models. Our findings suggest that activation-independent intrinsic mechanical properties of muscle are sufficient to stabilize contractions even in the dip region, offering new insight into how muscles maintain functional integrity during active stretch.
\end{abstract}

\begin{keywords}{Hill-type muscle model, simulation, dynamical instability, eigenvalue analysis, nondimensionalization.}\end{keywords}

\begin{MSCcodes}{34D20, 65L07, 74H55, 92C30.} \end{MSCcodes}

\section{Introduction}\label{sec1}

The sliding filament theory of muscle contraction (Hanson \& Huxley \cite{HansonHuxley1953}, Huxley \& Niedergerke \cite{HuxleyNiedergerke1954}, Huxley \cite{Huxley1957}) is a well-established mechanistic theory that relates the active force that is developed between the actin and myosin myofilaments within a sarcomere to the degree of overlap of those myofilaments and hence the sarcomere length. Cross-bridges between the actin and myosin are considered as independent force-generating units, and the number of cross-bridges that form depends on the degree of overlap between these myofilaments. At optimal length there is a degree of overlap that would allow the greatest number of cross-bridges to form. At  lengths greater than this optimal, the overlap between the actin and myosin decreases and thus the active force that can be developed by each sarcomere also decreases: this results in a negative slope to the sarcomere force-length relationship at longer lengths.

\begin{figure}
    \centering
    \begin{tabular}{cc}
        \subcaptionbox{\label{fig:ascending_descending_limbs_a}}{\includegraphics[width=0.46\textwidth]{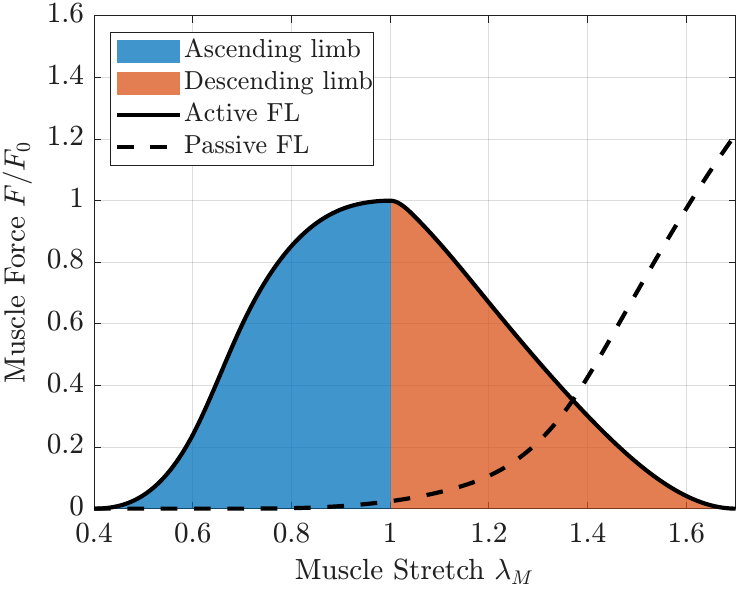}} &
        \subcaptionbox{\label{fig:ascending_descending_limbs_b}}{\includegraphics[width=0.46\textwidth]{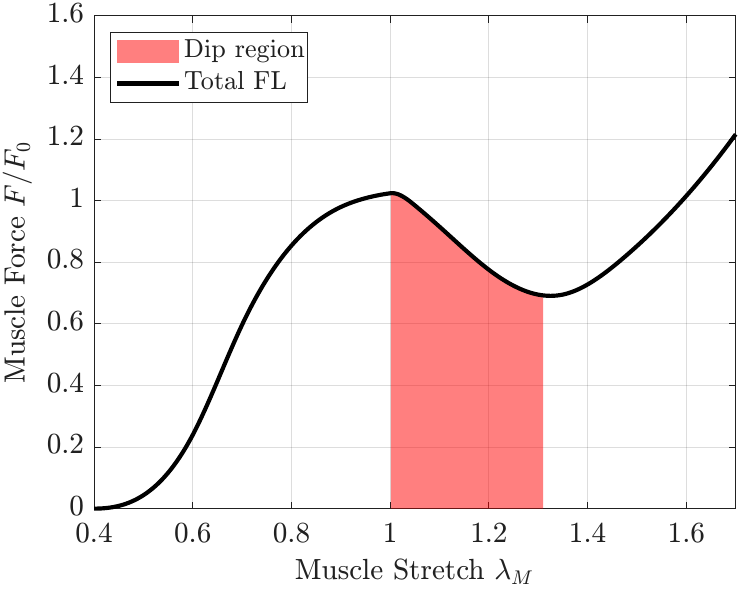}}
    \end{tabular}
    \caption{(a) Active and passive force-length (FL) relationships. The region $\lambda_M < 1$ corresponds to the ascending limb (in blue) since the curve has a positive slope. In turn, the region $\lambda_M > 1$ corresponds to the descending limb (in ochre) which has a distinctive negative slope. (b) When active and passive forces are added, they form the \textit{total} FL relationship, which typically has a dip region (i.e. a region with negative slope) around $1 < \lambda_M < 1.35$.}
\end{figure}

The negative slope of the active force-length relationship has also been described with empirical studies of muscle (Gordon et al. \cite{GordonHuxleyJulian1966}): the active force that muscles produce is minimal at very short lengths, has an “ascending limb” where the force increases at progressively longer lengths up to a plateau region that includes the optimal length, followed by a {\it ``descending limb''} where the active force decreases down to zero at progressively longer lengths with a negative slope (Fig. \ref{fig:ascending_descending_limbs_a}). When combined with the force-velocity relationship (Hill \cite{Hill1938}), these empirical descriptions form the basis of Hill-type muscle models (Zajac \cite{Zajac1989}) that are computationally efficient and have become the broadly used muscle models for applications in biomechanics (Wakeling et al. \cite{Wakeling2023Review}) and computer animation (Yeo et al. \cite{YeoEtAl2023NumericalInstability}).

Skeletal muscle would be considered dynamically unstable based on the negative slope (negative stiffness) of the descending limb of its active force-length relationship. There has been debate for many years over whether skeletal muscle is dynamically stable. {\it Dynamic stability} describes a situation where a system will return to its initial state following a small perturbation (Strogratz \cite{strogratz}). Using this definition, In 1953 A.V. Hill described the descending limb instability (DSI) for muscle lengths on the descending limb of the force-length relationship. A consequence of the negative slope at longer lengths leads to potential instability in the descending limb where a small perturbation in the lengths of adjacent in-series sarcomeres would result in a shorter and stronger sarcomere being adjacent to a longer but weaker sarcomere: the shorter sarcomere would thus tend to shorten further, stretching its weaker neighbour leading to a divergence in sarcomere lengths and an unstable situation (Hill \cite{Hill1953}). By contrast, on the ascending limb a small perturbation to the lengths of adjacent sarcomeres would lead to a longer and stronger sarcomere being adjacent to a shorter and weaker sarcomere: the longer sarcomere would tend to contract and shorten, stretching its shorter neighbour thus tending to restore the sarcomeres to a homogeneous length making the ascending limb dynamically stable. The general active force-length relationship is consistent across scales, from whole muscle (Winters \cite{Winters2011}), fibre (Edman \& Reggiani \cite{EdmanReggiani1987}), and sarcomere sizes (Moo \& Herzog \cite{MooHerzog2020}) and hence it is commonly assumed that the descending limb instability would be a phenomenon of active muscle across scales.

Energy-based mathematical descriptions were used to demonstrate that the active force-length relationship would be dynamically unstable on the descending limb for two adjacent sarcomeres that had a negative stiffness in that region (modelled as linear springs: Allinger et al. \cite{AllingerEpsteinHerzog1996}; Zahalak \cite{Zahalak1997}), and it was noted that the addition of sufficiently stiff components (with positive stiffness) could always stabilize the fibre (Zahalak \cite{Zahalak1997}). A one-dimensional multi-body model with 100 in-series Hill-type muscle models was used to visually demonstrate that small perturbations to sarcomere lengths would stabilize at lengths shorter (85\%) of the optimum length, but would diverge to a bimodal set on the dip region (Fig. \ref{fig:ascending_descending_limbs_b}) of the force-length curve (115\% of optimal length: Yeo et al. \cite{YeoEtAl2023NumericalInstability}). 

Actual muscle forces contain both active and passive components, with the passive component being strongly influenced by parallel elasticity from the protein titin at the sarcomere level (Granzier \& Labeit \cite{GranzierLabeit2007}) and additionally by the extracellular matrix at fibre-bundle through to whole muscle scales (Lieber \& Binder-Markey \cite{LieberBinderMarkey2021}). The passive force-length relationship for muscle is convex, rising to high forces at the longest length (Fig. \ref{fig:ascending_descending_limbs_a}). The total force-length relationship of a muscle has typically been considered as the sum of these active and passive components in a Hill-type muscle model (Zajac \cite{Zajac1989}), and thus has positive slopes and is considered stable on both its ascending limb, and at its longest lengths; however, there is still a “dip” region of the total force-length relationship with a negative slope that would be considered to be potentially unstable (Fig. \ref{fig:ascending_descending_limbs_b}). The dip region is particularly problematic if the goal of a muscle model is to simulate forces for active stretch of a muscle, or even isometric contractions using multi-body models (such as Yeo et al. \cite{YeoEtAl2023NumericalInstability}; G\"{u}nther et al. \cite{Gunther2012}; Ross \& Wakeling \cite{RossWakeling2016Multibody}). 
However, it should be noted that muscle is inherently stable and is built to withstand a life-time of use. Over the last few decades, considerable work has shown that the interactions between actin and titin molecules within each half-sarcomere result in greater resistance to stretch when the muscle is active in a process known as residual force enhancement (RFE) (Abbott \& Aubert \cite{AbbottAubert1952}; Herzog et al. \cite{HerzogLeeRassier2006}). This force-enhancement provides positive stiffness to the total force-length relationship, overcomes the descending limb instability (DLI) and has been observed across muscle scales from myofibrils and sarcomeres (Joumaa et al. \cite{JoumaaLeonardHerzog2008}; Leonard et al. \cite{LeonardDuvallHerzog2010}), isolated fibres (Edman et al. \cite{EdmanElzingaNoble1978}; Edman et al. \cite{EdmanElzingaNoble1982}) to whole muscles (Abbott \& Aubert \cite{AbbottAubert1952}; Herzog \& Leonard \cite{HerzogLeonard2002}). The effect of the actin-titin interactions was modelled by Heidlauf and co-workers [\cite{HeidlaufEtAl2017}] who embedded a cross-bridge model of the contractile elements into a three-dimensional continuum model of muscle. In this study the passive material properties of the muscle tissue were scaled down so that the muscle had a dip region in its static total-stress-stretch relationship, but additional positive stiffness was added using a “sticky spring” paradigm (Rode et al. \cite{RodeSiebertBlickhan2009}; Heidlauf et al. \cite{HeidlaufEtAl2016}) to represent actin-titin interactions during active stretch. This model showed that the actin-titin interactions convexified the total force-length relationship. This convexification stabilized sarcomere lengths during fixed-length and active stretches. Sarcomere lengths in this dip region of the force-length relationship converged and stabilized when these actin-titin interactions were present during muscle activity, but diverged without them (Heidlauf et al. \cite{HeidlaufEtAl2017}).
Despite the many decades of research into this problem, there is a fundamental question that remains to be addressed: are the material properties of muscle tissue sufficient to stabilize muscle contractions across its range of lengths? The force-length relationship that has been used to consider the stability question is inherently one-dimensional. Muscle (across the range of scales) is considered as a force actuator that develops force between two points: the muscle is implicitly considered to have no mass, volume, or shape beyond its length and thus it is one-dimensional. However, in reality muscle is a three-dimensional tissue that can undergo deformation in all three directions, the material properties of the tissue provide stiffness in all three directions, the tendency for the tissue to maintain its volume during deformation is its volumetric property, and tissue-mass can further damp down rapid accelerations during dynamic contractions. All these effects are beyond the scope of traditional 1D Hill-type models, but they may provide stability to the muscle tissue.

The purpose of this study is to consider whether activation-independent tissue properties can convexify the total force-length relationship of the muscle and provide dynamic stability at lengths longer than the optimal length. A more precise description of descending limb phenomena is made using tools from dynamical systems, and we demonstrate that the mechanical properties of the muscle tissue are sufficient to provide dynamical stability to the dip-region of a muscle's force-length relationship, and {this would occur across a range of scales from the sarcomere to the whole-muscle level.}

\section{Instability in a class of musculoskeletal models}\label{sec:unstable}

\begin{figure}
    \centering
    \includegraphics[width=0.9\textwidth]{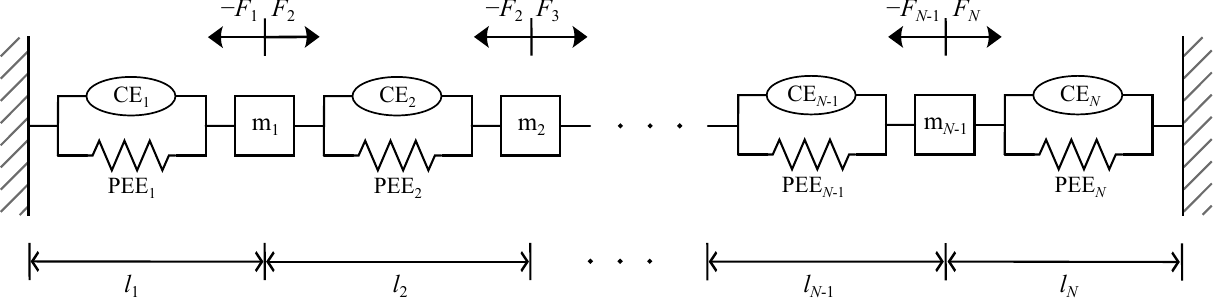}
    \caption{Mass enhanced multi-body muscle model. The muscle is divided into \textit{N} in-series segment, each containing a contractile element (CE), a parallel elastic element (PEE), and a point mass m.\label{fig:mass_enhanced_model}}
\end{figure}
We focus our attention on musculoskeletal models that represent muscle as a 1D structure of mass $m$, made of $N$ contractile elements connected in series via masses $m_i=m/(N-1)$, $i = 1, \dots, N-1$ (see Fig. \ref{fig:mass_enhanced_model}), similar to previous models introduced by Gunther et al. \cite{Gunther2012}, Ross et al. \cite{RossWakeling2016Multibody}, and Yeo et al. \cite{YeoEtAl2023NumericalInstability}, to name a few. We are particularly interested in the case of isometric contractions, and therefore the length of the muscle $L_M = \sum_{i=1}^N l_N$ is fixed at all times. Furthermore, we assume that the muscle length is given by $L_M = \lambda_M L_M^{opt}$, that is, the muscle is stretched by a multiple $\lambda_M$ from its optimal length $L_M^{opt}$. 
We assume that the length of each individual contractile unit, $l_i$, varies independently of other units. At time $t=0$, we have:
\begin{equation}
    l_i(0) = \lambda_M l_s^{opt},
\end{equation}
where $l_s^{opt} = L_M^{opt} / N$ is the optimal length of each segment. As the system departs from equilibrium, we have for $t > 0$:
\begin{equation}
    l_i(t) = l_i(0) + u_i(t), \quad i=1,\dots,N,
\end{equation}
where $u_i$ is the displacement of each contractile unit. Consequently, we can define the \textit{stretch} $\lambda_i$ and \textit{strain rate} $\depsilon_i$ of each segment as:
\begin{subequations} \label{eq:def_stretches_stability}
   \begin{align}
        \lambda_1(t) &= \dfrac{u_1(t)}{l_s^{opt}} + \lambda_M, \\
        \lambda_i(t) &= \dfrac{u_i(t)-u_{i-1}(t)}{l_s^{opt}} + \lambda_M, \quad i=2,\dots,N-1, \\
        \lambda_N(t) &= \dfrac{-u_{N-1}(t)}{l_s^{opt}} + \lambda_M,
   \end{align}
\end{subequations}
and,
\begin{equation} \label{eq:def_strain_rates_stability}
    \depsilon_i(t) = \dfrac{1}{\depsilon_0} \dfrac{d\lambda_i}{dt}, \quad i=1,\dots,N,
\end{equation}
where $\depsilon_0$ is the maximum strain rate of muscle. In this context, the dynamics of the model shown in Fig. \ref{fig:mass_enhanced_model} are given by the following set of second-order ordinary differential equations (ODEs):
\begin{subequations} \label{eq:odes_hill_only}
    \begin{align}
        m_1 \ddot{u}_1 &= F(\lambda_2, \depsilon_2) - F(\lambda_1, \depsilon_1), \\
        m_2 \ddot{u}_2 &= F(\lambda_3,\depsilon_3) - F(\lambda_2,\depsilon_2), \\
        &\hspace{0.6em} \vdots \notag \\
        m_{N-1} \ddot{u}_{N-1} &= F(\lambda_N, \depsilon_N) - F(\lambda_{N-1}, \depsilon_{N-1}),
    \end{align}
\end{subequations}
with initial conditions
\begin{equation}
    u_i(0) = \dot{u}_i(0) = 0, \quad i = 1,2,\dots, N-1.
\end{equation}
Here, each contractile unit generates force according to Hill's model:
\begin{equation} \label{eq:Hill_force_stability}
    F_i \equiv F(\lambda_i, \depsilon_i) := F_0 \left\{ F_A(\lambda_i) F_V(\depsilon_i) + F_P(\lambda_i) \right\}, \quad i=1,\dots,N,
\end{equation}
where $F_A$ and $F_P$ are respectively the active and passive force-length relationships (see Fig. \ref{fig:ascending_descending_limbs_a}), and $F_V$ is the force-velocity relationship of muscle.

\subsection{A numerical experiment} \label{sec:a_numerical_experiment}

\begin{figure}
    \centering
    \hspace*{-1.9em}\includegraphics[width=0.95\textwidth]{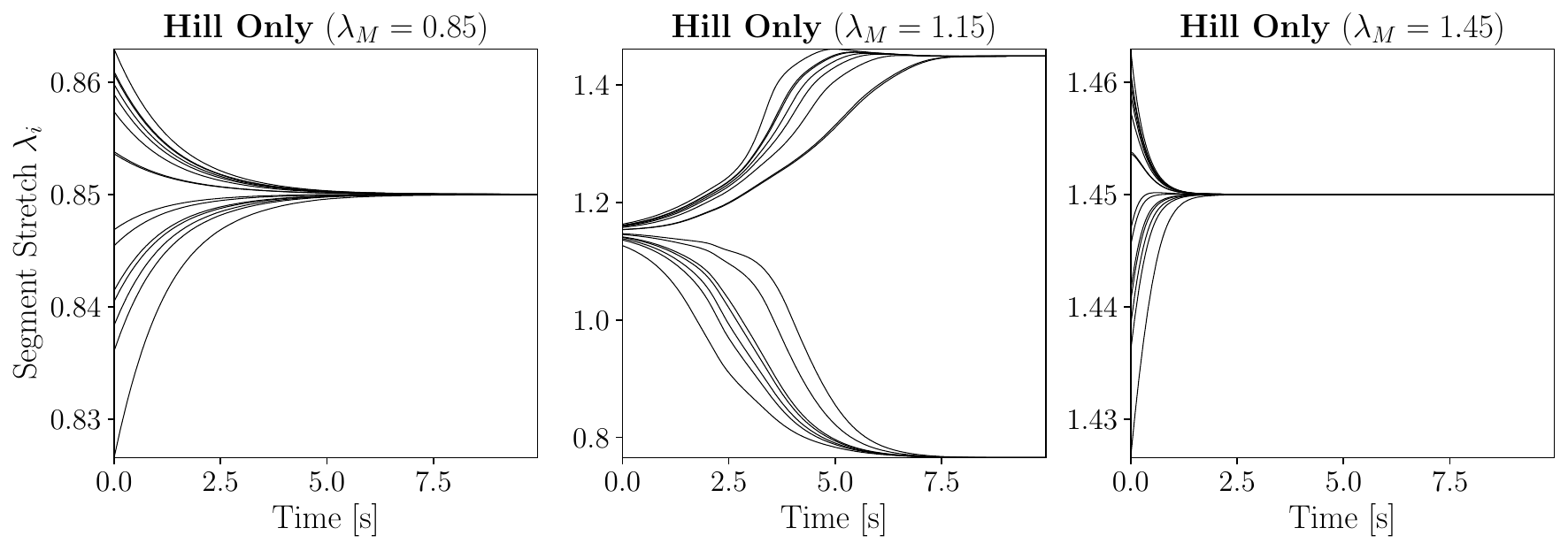}
    \caption{Stability of an isometric contraction according to the system of ODEs \eqref{eq:odes_hill_only}. At the beginning of the simulation, the contractile units (segments) are nonuniformly stretched at around 85\% (left), 115\% (middle), and 145\% (right) of the muscle's optimal length. At an initial stretch of 115\% ($\lambda_M = 1.15$), the descending limb instability causes the segments to bifurcate to two different lengths.}
    \label{fig:hill_only_unstable}
\end{figure}

The lack of \textit{dynamical} stability in this type of system has recently been shown by Yeo et al. \cite{YeoEtAl2023NumericalInstability} using a numerical experiment, which we replicate here for completeness. We consider a muscle of length $L_M^{opt} = 0.3$ m with maximum isometric force $F_0 = 525 $ N. Without loss of generality, we set the maximum strain rate to $\depsilon_0 = 5$ s$^{-1}$. We also take $N=17$ masses and an initial condition $u_i(0) = \eta_i$, $\dot{u}_i(0) = 0$, for $i=1,\dots,N-1$. Here  $\eta_i$ is a random variable drawn from a uniform distribution on $[-0.01, 0.01]$. Then, we consider isometric contractions for three different whole-muscle stretches: $\lambda_M = 0.85$, $1.15$, and $1.45$. 
As expected, the system is dynamically stable whenever,
\begin{equation} \label{eq:positive_derivative}
    \pder{F}{\lambda}(\lambda_M, 0) > 0,
\end{equation}
and unstable when this condition fails (for example, when $\lambda_M  = 1.15$, see Fig. \ref{fig:hill_only_unstable}). This can also be verified analytically through an eigenvalue analysis of the linearized problem (see Appendix \ref{secA1}). The analysis differs from that of Zahalak [\cite{Zahalak1997}] in the sense that the force $F(\lambda,\depsilon)$ in \eqref{eq:positive_derivative} does not necesarily have to follow a linear relationship (e.g., Hooke's law).

\subsection{The continuum limit}

As our stabilization strategy will be based on arguments from continuum mechanics (more precisely, those found in 3D models), we briefly turn our attention to the continuum limit of the system of ODEs \eqref{eq:odes_hill_only} as $N \to \infty$. To simplify matters, consider the model shown in Fig. \ref{fig:mass_enhanced_model} where the fixed right end is replaced by a moving mass $m_N$ which is pulled with a constant force $F_{ext}$. Furthermore, assume that $m_i = m/N$, $i=1,\dots,N$, and that the muscle is not pre-stretched (i.e. $\lambda_M = 1$). In addition, define $l_s^{opt} = L_M^{opt}/N =: \Delta x$. After some algebraic manipulations, the system \eqref{eq:odes_hill_only} becomes:
\begin{subequations} \label{eq:continuum_model_1}
    \begin{align}
        \dfrac{m}{L_M^{opt}} \ddot{u}_1 &= \dfrac{F(\lambda_2, \depsilon_2) - F(\lambda_1, \depsilon_1)}{\Delta x}, \\
        \dfrac{m}{L_M^{opt}} \ddot{u}_2 &= \dfrac{F(\lambda_3,\depsilon_3) - F(\lambda_2,\depsilon_2)}{\Delta x}, \\
        &\hspace{0.6em} \vdots \notag \\
        \dfrac{m}{L_M^{opt}} \ddot{u}_{N} &= \dfrac{F_{ext} - F(\lambda_{N}, \depsilon_{N})}{\Delta x}.
    \end{align}
\end{subequations}
In particular, for $i=1,\dots,N$, the segment stretches $\lambda_i$ and strain rates $\depsilon_i$ now read:
\begin{equation} \label{eq:continuum_model_2}
    \lambda_i = 1 + \dfrac{u_i-u_{i-1}}{\Delta x},
\end{equation}
and
\begin{equation} \label{eq:continuum_model_3}
    \depsilon_i = \dfrac{1}{\depsilon_0} \dfrac{d}{dt} \left( \dfrac{u_i-u_{i-1}}{\Delta x} \right).
\end{equation}

We are interested in studying the behaviour of the system \eqref{eq:continuum_model_1}-\eqref{eq:continuum_model_3} as $N \to \infty$, or equivalently, as $\Delta x \to 0$. This is the \textit{continuum limit}. In this case, equations \eqref{eq:continuum_model_2} and \eqref{eq:continuum_model_3} define respectively the following the quantities:
\begin{equation}
    \lambda(x,t) := \lim_{\Delta x \to 0} \lambda_i = 1 + \pder{u}{x},
\end{equation}
and
\begin{equation}
    \depsilon(x,t) = \lim_{\Delta x \to 0} \depsilon_i = \dfrac{1}{\depsilon_0} \pder{}{t}\left(\pder{u}{x} \right) = \dfrac{1}{\depsilon_0} \pder{\lambda}{t}.
\end{equation}
That is, we have recovered the one-dimensional version of the stretch and strain rate variables that are commonly considered in solid mechanics. Moreover, we can make the following claim with respect to the system \eqref{eq:continuum_model_1}.

\begin{claim} \label{cl:hill_pde}
The system of ODEs \eqref{eq:continuum_model_1} corresponds to a first-order method of lines discretization of the partial differential equation (PDE):
\begin{equation} \label{eq:hill_pde_1}
    \rho_{0,L} \pder{^2 u(x,t)}{t^2}  = \pder{F(\lambda(x,t),\depsilon(x,t))}{x}, \quad 0 < x < L_M^{opt}, \quad 0 < t < \infty,
\end{equation}
subject to boundary conditions
\begin{subequations} \label{eq:hill_pde_2}
    \begin{align}
        u(0,t) &= 0, \\
        F(\lambda(\cdot, t),\depsilon (\cdot, t))\Big|_{x\to (L_M^{opt})^+} &= F_{ext}(t), \label{eq:Neumann_F_ext}
    \end{align}
\end{subequations}
and initial conditions
\begin{equation} \label{eq:hill_pde_3}
    u(x,0) = \pder{u}{t} = 0.
\end{equation}
Here, $F(\lambda,\depsilon)$ corresponds to Hill's force defined in \eqref{eq:Hill_force_stability} and $\rho_{0,L} := m/L_M^{opt}$ is a linear density.
\end{claim}

Note that the boundary condition \eqref{eq:Neumann_F_ext} is nonlinear and depends on $\frac{\partial u}{\partial x}$, hence, it can be viewed as a ``generalized Robin'' condition. Thus, the problem \eqref{eq:hill_pde_1}-\eqref{eq:hill_pde_3} is at least well-defined. Whether it is \textit{well-posed} (that is, given a smooth $F_{ext}$, the problem has a unique regular solution depending continuously on $F_{ext}$) is a much more difficult question to answer. As it is known, this can only be done in limited scenarios (for instance, when $F$ is linear, see Truesdell \& Noll \cite{TruesdellNoll2004}), so a general result cannot be stated for arbitrary $F$ and $F_{ext}$.

\begin{figure}
    \centering
    \includegraphics[width=0.95\linewidth]{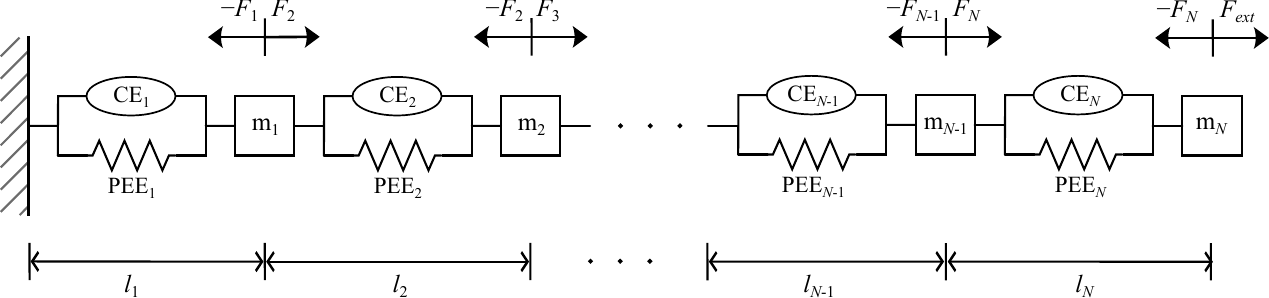}
    \caption{Method of lines discretization of the PDE \eqref{eq:hill_pde_1} using $N$ masses.}
    \label{fig:mol_pde}
\end{figure}

\begin{proof}[Proof of Claim 1] 
    Consider a discretization $\{x_0, x_1, \dots, x_N\}$ with constant spacing $\Delta x := L_M^{opt}/N$ of the interval $[0,L_M^{opt}]$ where $x_0 = 0$ and $x_N = L_M^{opt}$, then, define $u_i := u(x_i,t)$ for $t \geq 0$, $i=1,\dots,N$. Furthermore, consider a backwards first-order formula for the stretch and strain rates, that is,
    \[
        \lambda_i = \lambda(x_i,t) = 1 + \dfrac{u_i-u_{i-1}}{\Delta x},
    \]
    and
    \[
        \depsilon_i = \depsilon(x_i, t) = \dfrac{1}{\depsilon_0}\dfrac{d\lambda_i}{dt}.
    \]
    These are precisely \eqref{eq:continuum_model_2} and \eqref{eq:continuum_model_3}. Now, to discretize the spatial derivative on the right-hand side of \eqref{eq:hill_pde_1}, consider a forwards first-order difference at $x_1$, \dots, $x_{N}$:
    \begin{equation*} 
        \rho_{0,L} \dfrac{d^2 u_i}{dt^2} = \dfrac{F(\lambda_{i+1},\depsilon_{i+1}) - F(\lambda_i, \depsilon_i)}{\Delta x}, \quad i =1, \dots, N.
    \end{equation*}
    We show this discretization in Fig. \ref{fig:mol_pde}. Given that we cannot compute $F(\lambda_{N+1}, \depsilon_{N+1})$, we replace this quantity by $F_{ext}$, based on the boundary condition \eqref{eq:Neumann_F_ext}. Thus, we have recovered the set of equations \eqref{eq:continuum_model_1}
\end{proof}

The PDE in \eqref{eq:hill_pde_1} is also a model that allows us to discard other factors that could cause instability, such as mass, maximum strain rate, or optimal muscle length. First, let:
\begin{equation}
    \widehat{F} = \dfrac{F}{F_0}, 
\end{equation}
denote the nondimensional version of the Hill force term. We also define some nondimensional variants of the displacement, space, and time variables:
\begin{equation}
    \widehat{u} = \dfrac{u}{L_M^{opt}}, \quad \widehat{x} = \dfrac{x}{L_M^{opt}}, \quad \widehat{t} = \dfrac{t}{1/\depsilon_0}.
\end{equation}
Plugging these quantities into \eqref{eq:hill_pde_1} we obtain:
\begin{equation} \label{eq:nondim_1}
    \dfrac{\rho_{0,L} (L_M^{opt})^2 \depsilon_0^2}{F_0} \, \pder{^2 \widehat{u}}{\widehat{t}^2} = \pder{\widehat{F}}{\widehat{x}}.
\end{equation}
All terms in the previous equation are nondimensional, but we can further rearrange the terms in the left-hand side to obtain a more physically meaningful number. 

Recall that $\rho_{0,L}$ is a \textit{linear} density (it has units of $\text{kg} \ \text{m}^{-1}$). We can convert this quantity into the more common \textit{volumetric} density $\rho_0$ through a representative cross-sectional area (CSA), that is,
\begin{equation}
    \rho_{0,L} = \rho_0 \cdot CSA.
\end{equation}
Moreover, because the maximum isometric force $F_0$ can be written in terms of the maximum isometric stress as $F_0 = \sigma_0 \cdot CSA$, the number on the left-hand side of \eqref{eq:nondim_1} can be written as
\begin{equation}
    \dfrac{\rho_0 (L_M^{opt})^2 \depsilon_0^2}{F_0} = \dfrac{\rho_0 (L_M^{opt})^2 \depsilon_0}{\sigma_0 / \depsilon_0}.
\end{equation}
The nondimensional number on the right-hand side of this expression can be viewed as a ratio between inertial forces (due to mass) and viscous forces. We call this number the \textit{musculoskeletal Reynolds number} ($\text{Re}_{MSK}$), akin to the Reynolds number commonly found in fluid dynamics. In this way, we can rewrite \eqref{eq:nondim_1} as:
\begin{equation} \label{eq:nondim_final}
    \text{Re}_{MSK} \pder{^2 \widehat{u}}{\widehat{t}^2} = \pder{\widehat{F}}{\widehat{x}}.
\end{equation}
The definition of $\text{Re}_{MSK}$ and the nondimensional equation \eqref{eq:nondim_final} suggest that the dynamic stability of the multi-body model \eqref{eq:odes_hill_only} does not depend on the size of the muscle (in terms of density $\rho_0$ and optimal muscle length $L_M^{opt}$) or the maximum unloaded strain rate $\depsilon_0$, since at equilibrium, the left-hand side vanishes.

\section{Stabilization}

The descending limb instability (DLI), caused by the negative stiffness of the total force-length curve (i.e. failure to meet condition \eqref{eq:positive_derivative}) can be overcome by adding \textit{any} stiff-enough parallel elastic element to Hill's model \eqref{eq:Hill_force_stability}. As an alternative to activation-dependent elements (such as titin), we can consider adding a passive base material component, inspired by those found in 3D models (see, e.g., Almonacid et al. \cite{AlmonacidEtAl2022_SIAP_Paper}). In particular, let us assume this base material component behaves as Neo-Hookean material:
\begin{equation} \label{eq:force_neohookean}
    F_{NH}(\lambda) = \left\{
        \begin{aligned}
            &F_0 \dfrac{\kappa}{\sigma_0} \left( \widehat{\mathcal{L}}_{\mu} \lambda +  \dfrac{\widehat{\mathcal{L}}_{\lambda} \log(\lambda) - \widehat{\mathcal{L}}_\mu}{\lambda}\right), \quad &\lambda \geq 1, \\
            &0, \quad &\text{otherwise}.
        \end{aligned}
    \right.
\end{equation}
Here, $\kappa$ is the bulk modulus of muscle, $\sigma_0$ is the maximum isometric stress, and the (non-dimensional) Lam\'{e} parameters are given by
\begin{equation}
    \widehat{\mathcal{L}}_{\lambda} = \dfrac{\nu}{(1-\nu)(1-2\nu)}, \quad \widehat{\mathcal{L}}_{\mu} = \dfrac{1}{2(1+\nu)},
\end{equation}
with $\nu$ being the Poisson's ratio of muscle tissue, which can be set to $\nu = 0.499$ (see Kuthe \& Uddanwadiker \cite{Kuthe2016}) owing to the (near-)incompressibility of muscle.
In this way, we consider the following expression for the muscle force in \eqref{eq:hill_pde_1}:
\begin{equation} \label{eq:force_hill_plus_bm}
    F(\lambda, \depsilon) = (1-\chi_{BM}) F_{Hill}(\lambda, \depsilon) + \chi_{BM} F_{NH}(\lambda),
\end{equation}
where $\chi_{BM}$ denotes the fraction of base material that could be present in a muscle fibre. The dynamics of this 1-D system (in nondimensional form) are described by the PDE:
\begin{equation} \label{eq:nondim_pde_hill_nh}
    \text{Re}_{MSK} \pder{^2 \widehat{u}}{\widehat{t}^2} = (1-\chi_{BM})\, \pder{\widehat{F}_{Hill}(\lambda,\depsilon)}{\widehat{x}} + \chi_{BM} \,  \dfrac{\kappa}{\sigma_0}  \, \pder{\widehat{F}_{NH}(\lambda)}{\widehat{x}}.
\end{equation}
Consequently, the stability of the system now depends on the fraction of base material $\chi_{BM}$, as well as the bulk modulus to maximum isometric stress ratio $\kappa / \sigma_0$.


Following the result in Claim \ref{cl:hill_pde}, we can discretize the PDE \eqref{eq:nondim_pde_hill_nh} using the method of lines to derive a stabilized version of \eqref{eq:odes_hill_only} for an isometric contraction: 
\begin{multline} \label{eq:odes_hill_stabilized}
    \quad m_{i-1} \ddot{u}_{i-1} =
    (1-\chi_{BM})\left[ F_{Hill}(\lambda_i, \depsilon_i) - F_{Hill}(\lambda_{i-1}, \depsilon_{i-1}) \right] \\ 
    + \chi_{BM}\left[ F_{NH}(\lambda_i) - F_{NH}(\lambda_{i-1}) \right], \quad i = 1,\dots,N-1, \quad 
\end{multline}
with the stretches $\lambda_i$ and strain rates $\depsilon_i$ as in \eqref{eq:def_stretches_stability} and \eqref{eq:def_strain_rates_stability}, respectively. 

\begin{figure}
    \centering
    \begin{tabular}{cc}
        \subcaptionbox{\label{fig:forces_stabilized_a}}{\includegraphics[width=0.45\textwidth]{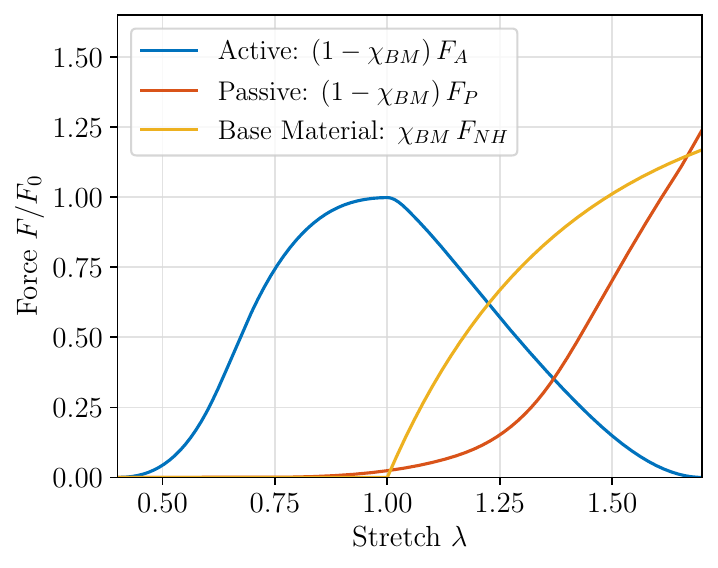}} &
        \subcaptionbox{\label{fig:forces_stabilized_b}}{\includegraphics[width=0.45\textwidth]{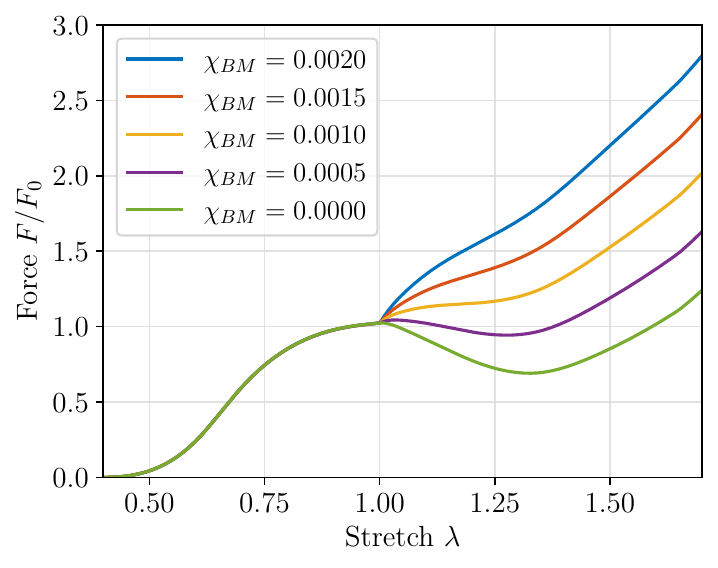}}
    \end{tabular}
    \caption{(a) Forces in the stabilized model ($\chi_{BM} = 0.0015$). (b) Normalized total force for different values of $\chi_{BM}$.}
\end{figure}

We can repeat the experiment from Section \ref{sec:a_numerical_experiment} to test the stability of the system \eqref{eq:odes_hill_stabilized}. In particular, we take in equation \eqref{eq:force_neohookean} a bulk modulus $\kappa = 10^6 \ \text{Pa}$ and a maximum isometric stress $\sigma_0 = 200 \ \text{kPa}$, which are values commonly assumed in 3D models (see, e.g.,  Almonacid et al. \cite{AlmonacidEtAl2022_SIAP_Paper}). Furthermore, we consider $\chi_{BM} = 0.0015$, that is, we assume that 99.85\% of the force that a muscle fibre produces is due to Hill's force and 0.15\% from a base material contribution. The resulting active, passive, and base material forces are shown in Fig. \ref{fig:forces_stabilized_a}. Smaller values of $\chi_{BM}$ can also yield a total FL curve that lacks a dip region, as shown in Fig. \ref{fig:forces_stabilized_b}.


\begin{figure}
    \centering
    \hspace*{-1.9em}\includegraphics[width=0.95\textwidth]{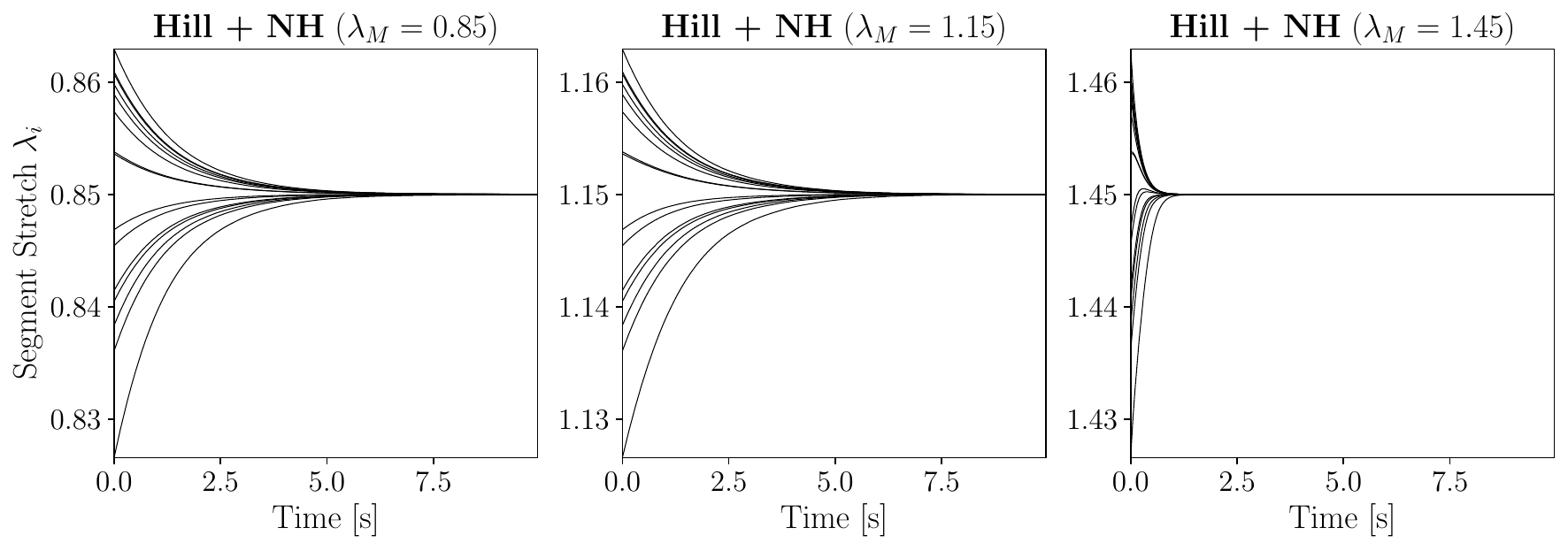}
    \caption{Stability of an isometric contraction according to the system of ODEs \eqref{eq:odes_hill_stabilized}. As before, the elements (as well as the muscle) are stretched initially at around 85\% (left), 115\% (middle), and 145\% (right) of the muscle's optimal length. In all cases, the contractions are stable since the derivative of the total force (made of active, passive, and base material components) is always positive. The stability conferred by the base material properties in the dip-region of the force-length relationship (stretch of 1.15) should be contrasted with the instability shown in Fig.3}
    \label{fig:hill_plus_nh_stable}
\end{figure}

\begin{figure}
    \centering
    \includegraphics[width=0.99\linewidth]{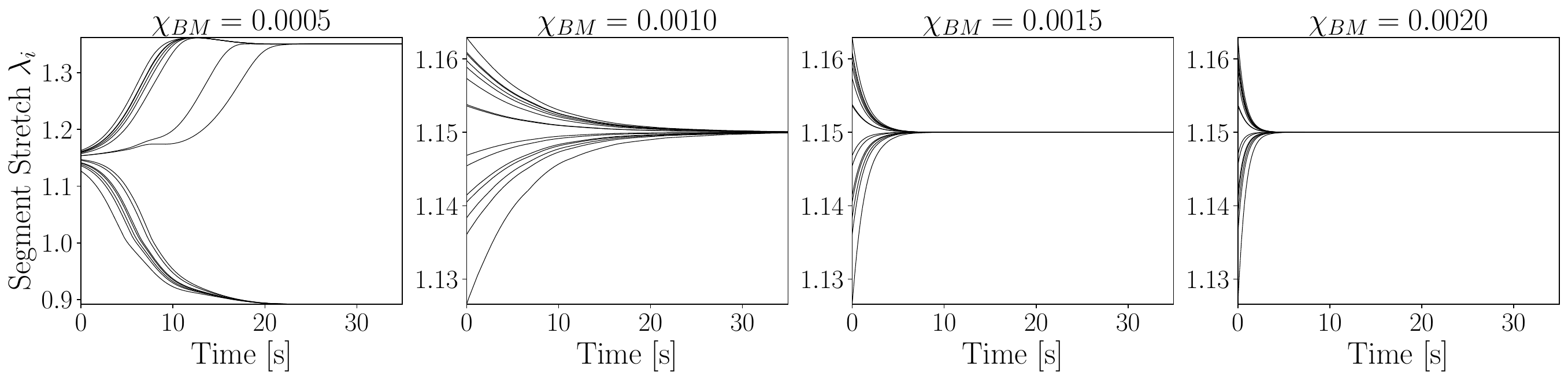}
    \caption{Stabilization along the descending limb ($\lambda_M = 1.15)$ using different values of the base material fraction $\chi_{BM}$. Note that the contraction appears to be stable whenever the value of $\chi_{BM}$ yields a force-length relationship without a dip region (see Fig. \ref{fig:forces_stabilized_b}).}
    \label{fig:different_bm}
\end{figure}

For a base material fraction $\chi_{BM} = 0.0015$, we show in Fig. \ref{fig:hill_plus_nh_stable} the results for the same three isometric contractions as in Section \ref{sec:a_numerical_experiment}. We now observe (1): no difference for $\lambda_0 = 0.85$ since the Neo-Hookean term does not act for stretches below 1, (2): stabilization on the descending limb of the active FL curve, shown in the figure for $\lambda_0 = 1.15$, and (3): rapid stabilization due to a stiffer passive element. Furthermore, we show in Fig. \ref{fig:different_bm} the effect of the base material fraction in the contraction dynamics on the descending limb ($\lambda_M = 1.15$).


\section{Summary}

\begin{enumerate}
    \item We used a one-dimensional multi-body model of muscle (Fig. \ref{fig:mass_enhanced_model}) to demonstrate that the dip-region of the total force length curve (Fig. \ref{fig:ascending_descending_limbs_b}) is dynamically unstable. These results supported a previous report on this issue (Yeo et al. \cite{YeoEtAl2023NumericalInstability}). This dynamical instability occurred when the stretches of adjacent muscle segments were given small and random perturbations, and this was followed by the stretches of these muscle segments bifurcating to a bimodal distribution (Fig. 3). These findings concur with the predictions initially made by A.V. Hill in [\cite{Hill1953}].
    \item This dynamical instability persists in a continuum formulation of the problem, where the segment lengths approach the limit of being infinitesimally small. Thus, the case of dynamical instability for these 1D models would still occur down to a sarcomere-length scale.
    \item We introduced non-dimensional variants of displacement, space and time to identify a ratio between the inertial forces (due to mass) and viscous forces that we term the musculoskeletal Reynolds number ($\text{Re}_{MSK}$). The definition of this $\text{Re}_{MSK}$ and the non-dimensional equation \eqref{eq:nondim_final} suggest that the dynamical stability and instability of the multi-body model \eqref{eq:odes_hill_only} does not depend on the size of the muscle or its maximum unloaded strain-rate.
    \item Previous models have shown how convexifying the total force-length curve for a muscle confers dynamical stability when the muscle is in the dip-region of its total force-length relationship. This has been demonstrated by including the activatable interaction between the proteins actin and titin (Heidlauf et al. \cite{HeidlaufEtAl2017}), and indeed it had previously been suggested that any sufficiently stiff component within the muscle could stabilize its force-length relationship (Zahalak \cite{Zahalak1997}).
    \item We note that muscle is three-dimensional with shape, volume and tissue properties. The volumetric and base-material tissue properties confer a positive stiffness to the muscle when it is stretched. We introduced these volumetric and tissue effects into the 1D multi-body model \eqref{eq:nondim_final} by adding additional stiffness \eqref{eq:force_neohookean} (as a Neo-Hookean material) and demonstrate that they also convexify the force-length relationship (Fig. \ref{fig:forces_stabilized_b}) of the muscle where the dip-region occurred (Fig. \ref{fig:ascending_descending_limbs_b}). Thus, 3D muscle tissue properties provide dynamical stability to the descending limb of force-length relationship of the muscle (Fig. \ref{fig:hill_plus_nh_stable}), including in its dip-region.
    \item These findings suggest that the inherent tissue properties of muscle keep it dynamically stable, even before the effects of activatable actin-titin interactions are considered. 
\end{enumerate}

\section{Acknowledgements}

We acknowledge funding from NSERC of Canada Discovery grants to N.N. and J.M.W. 

\subsection*{Data Availability}

The results in this work can be reproduced using the Python scripts available at \href{http://www.github.com/javieralmonacid/stable-muscle-fibres}{github.com/javieralmonacid/stable-muscle-fibres}.

\subsection*{Conflict of interest}

The authors declare that they have no conflict of interest.









\appendix
\section{Mathematical characterization of the DLI}\label{secA1}


Using tools from dynamical analysis, we can study the stability of the system of ODEs \eqref{eq:odes_hill_only} to provide a mathematical characterization of the DLI.
First, we rewrite \eqref{eq:odes_hill_only} as a system of first order ODEs, which we can compactly express as:
\begin{equation} \label{eq:odes_hill_only_first_order}
\dfrac{d}{dt} \begin{pmatrix}
    \vec{u} \\ \vec{v} 
\end{pmatrix} = \dfrac{N-1}{m}\begin{bmatrix}
    \*0 & \*I \\ \*0 & \*0
\end{bmatrix} \begin{pmatrix}
    \vec{u} \\ \vec{v} 
\end{pmatrix} + \mathcal{F}(\vec{u}, \vec{v}),
\end{equation}
where $\vec{u} = (u_1, \dots, u_{N-1})^\top$ is the vector of displacements, $\vec{v} = (v_1, \dots, v_{N-1})^\top$ is the vector of velocities (not to be confused with the strain rate variable defined in \eqref{eq:def_strain_rates_stability}), and $\mathcal{F}(\vec{u}, \vec{v})$ is a nonlinear term given by:
\begin{equation} \label{eq:def_nonlinear_force_term}
    \mathcal{F}(\vec{u}, \vec{v}) = \begin{pmatrix}
        \vec{0} \\ \vec{F}(\vec{u}, \vec{v})
    \end{pmatrix}, \quad 
    \vec{F}(\vec{u},\vec{v}) = \begin{pmatrix}
        F(\lambda_2, \depsilon_2) - F(\lambda_1, \depsilon_1) \\
        \vdots \\
        F(\lambda_N, \depsilon_N) - F(\lambda_{N-1}, \depsilon_{N-1})
    \end{pmatrix},
\end{equation}
with the force $F$ corresponding to Hill's force with maximal activation ($a=1$), that is
\begin{equation}
    F(\lambda, \depsilon) = F_0 \Big\{ F_A(\lambda)F_V(\depsilon) + F_P(\lambda) \Big\}.
\end{equation}

Notice that the zero solution, i.e. $(\vec{u}, \vec{v}) = (\vec{0}, \vec{0})$, is an equilibrium point of the system \eqref{eq:odes_hill_only_first_order}. 
Therefore, according to the theorem of Hartman-Grobman (see, e.g., Coayla-Teran et al. \cite{HartmanGrobman}), we may linearize \eqref{eq:odes_hill_only_first_order} around $(\vec{0}, \vec{0})$ to analyze the stability of this equilibrium point. Because changes in stretch variable will be small relative to the time it takes for the system to stabilize, we can assume that $\depsilon_i \approx 0$. Thus, it suffices to analyze the linear system:
\begin{equation}
    \dfrac{d}{dt} \begin{pmatrix}
        \vec{u} \\ \vec{v} 
    \end{pmatrix} = J(\vec{0}, \vec{0}) \begin{pmatrix}
        \vec{u} \\ \vec{v}
    \end{pmatrix},
\end{equation}
where the Jacobian $J(\vec{u}, \vec{v})$, for arbitrary $(\vec{u},\vec{v})$, is given by
\begin{equation}
    J(\vec{u}, \vec{v}) = \begin{bmatrix}
        \*0 & \frac{N-1}{m} \*I_{N-1} \\
        \frac{\partial \vec{F}}{\partial \vec{u}} & \*0
    \end{bmatrix}.
\end{equation}
Here, $\*I_{N-1} \in \R^{N-1 \times N-1}$ is the identity matrix and $\frac{\partial \vec{F}}{\partial \vec{u}}$ the tridiagonal matrix of entries:
\begin{equation}
    \left( \dfrac{\partial \vec{F}}{\partial \vec{u}} \right)_{i,j} = \left\{
    \begin{aligned}
        &-\frac{1}{l_s^{opt}} \left[ \pder{F}{\lambda}(\lambda_{i+1},0) + \pder{F}{\lambda}(\lambda_i, 0) \right] \quad &j = i, \\[0.5em]
        &\frac{1}{l_s^{opt}} \pder{F}{\lambda}(\lambda_i,0) \quad & j = i+1,\\[0.5em]
        &0 \quad &\text{otherwise}.
    \end{aligned}
    \right.
\end{equation}
In particular, when $(\vec{u},\vec{v}) = (\vec{0},\vec{0})$, we have 
\begin{equation}
    J(\vec{0}, \vec{0}) = \left[ \begin{array}{cc}
        \*0 & \frac{N-1}{m} \*I_{N-1} \\
        \frac{1}{l_s^{opt}}\frac{\partial F}{\partial \lambda}(\lambda_M,0) \, \*D_0 & \*0
    \end{array} \right],
\end{equation}
where $\*D_0 \in \R^{N-1 \times N-1}$ is the matrix
\begin{equation}
    \*D_0 = \left[ \begin{array}{ccccc}
        -2 & 1 & 0 & \cdots & 0 \\
        1 & -2 & 1 & \cdots & 0 \\
          & \ddots & \ddots & \ddots\\
        0 &\cdots & 0 & 1 & -2
    \end{array} \right].
\end{equation}
The latter is a matrix typically found in finite difference schemes for ODEs and PDEs (see, e.g., Strikwerda \cite{Strikwerda}). Its eigenvalues are:
\begin{equation}
    \nu_k = -4 \sin^2 \left( \dfrac{k\pi}{2N} \right), \quad k = 1, \dots, N-1.
\end{equation}
Therefore, to find the eigenvalues $\omega_k$ of $J(\vec{0},\vec{0})$, we may use the following identity:
\[
    \det \left(\omega \*I_{2(N-1)} - J(\vec{0}, \vec{0}) \right) = \det \left(\omega^2 \*I_{N-1} - \dfrac{N-1}{l_s^{opt} m} \dfrac{\partial F}{\partial \lambda}(\lambda_M, 0) \*D_0 \right),
\]
which yields two cases, depending on the sign of $\frac{\partial F}{\partial \lambda}(\lambda_M, 0)$:
\begin{enumerate}
    \item If $\frac{\partial F}{\partial \lambda}(\lambda_M, 0) > 0$ (that is, the muscle is contracted on the \textit{ascending} limb of the FL relationship), then the eigenvalues of the Jacobian are:
\begin{equation}
    \omega_k = \pm 2 \dfrac{N-1}{l_s^{opt} m} \frac{\partial F}{\partial \lambda}(\lambda_M, 0) \sin \left( \dfrac{k\pi}{2N} \right) \iota, \quad k = 1, \dots, N-1,
\end{equation}
with $\iota=\sqrt{-1}$. Although all the eigenvalues are purely complex, they are distinct, and therefore the system \eqref{eq:odes_hill_only_first_order} is stable according to Hairier et al. [\cite{HairerNorsettWanner}; Theorem 13.1].
\item If $\frac{\partial F}{\partial \lambda}(\lambda_M, 0) < 0$ (that is, the muscle is contracted on the dip region of the FL relationship), then the eigenvalues are purely real:
\begin{equation}
    \omega_k = \pm 2 \dfrac{N-1}{l_s^{opt} m} \frac{\partial F}{\partial \lambda}(\lambda_M, 0) \sin \left( \dfrac{k\pi}{2N} \right), \quad k = 1, \dots, N-1.
\end{equation}
Therefore, because of the presence of positive eigenvalues, the system \eqref{eq:odes_hill_only_first_order} is dynamically unstable. This is a mathematical characterization of the DLI.
\end{enumerate}





\bibliographystyle{plain}
\bibliography{sn-bibliography}

\end{document}